\documentclass[twocolumn]{aastex62}

\usepackage{longtable} 
\usepackage{graphicx}
\usepackage{subfigure}
\usepackage{multirow}

\usepackage{lineno}
\def \mgii  {Mg\,{\sc ii}}
\def \cii     {C\,{\sc ii}}

\def \siiv   	{Si\,{\sc iv}}

\def \six   	{Si\,{\sc x}}
\def \sx    	{S\,{\sc x}}

\def \feviii	{Fe\,{\sc viii}}
\def \feix		{Fe\,{\sc ix}}
\def \fex		{Fe\,{\sc x}}
\def \fexi		{Fe\,{\sc xi}}
\def \fexii		{Fe\,{\sc xii}}
\def \fexiv		{Fe\,{\sc xiv}}
\def \fexv		{Fe\,{\sc xv}}
\def \fexvi		{Fe\,{\sc xvi}}
\def \fexiii	{Fe\,{\sc xiii}}
\def \fexvi		{Fe\,{\sc xvi}}

\def \iris   	{{\it IRIS}}
\def \skylab   	{{\it Skylab}}
\def \muse   	{{\it MUSE}}
\def \sdo   	{{\it SDO}}
\def \hinode   	{{\it Hinode}}
\def \arcsec 	{\hbox{$^{\prime\prime}$}}

\def \alfven  {Alfv$\acute{\rm e}$n}

\received{}
\submitjournal{ApJ}

\shorttitle{Coronal Abundances and Underlying chromospheric/TR properties}
\shortauthors{Testa et al.}

\begin{document}

\title{Coronal Abundances in an Active Region: Evolution and Underlying Chromospheric and Transition Region Properties}
 
\correspondingauthor{Paola Testa}
\email{ptesta@cfa.harvard.edu}

\author[0000-0002-0405-0668]{Paola Testa}
\affil{Harvard-Smithsonian Center for Astrophysics,
60 Garden St, Cambridge, MA 02193, USA}

\author[0000-0002-0333-5717]{Juan Mart\'inez-Sykora}
\affil{Lockheed Martin Solar \& Astrophysics Laboratory,
3251 Hanover St, Palo Alto, CA 94304, USA}
\affil{Bay Area Environmental Research Institute, NASA Research Park, Moffett Field, CA 94035, USA.}
\affil{Rosseland Centre for Solar Physics, University of Oslo, P.O. Box 1029 Blindern, N-0315 Oslo, Norway}
\affil{Institute of Theoretical Astrophysics, University of Oslo, P.O. Box 1029 Blindern, N-0315 Oslo, Norway}

\author[0000-0002-8370-952X]{Bart De Pontieu}
\affil{Lockheed Martin Solar \& Astrophysics Laboratory,
3251 Hanover St, Palo Alto, CA 94304, USA}
\affil{Rosseland Centre for Solar Physics, University of Oslo, P.O. Box 1029 Blindern, N-0315 Oslo, Norway}
\affil{Institute of Theoretical Astrophysics, University of Oslo, P.O. Box 1029 Blindern, N-0315 Oslo, Norway}

%\author{others}
%\affiliation{}

% The abstract

\begin{abstract}
The element abundances in the solar corona and solar wind are often different from those of the solar photosphere, typically with a relative enrichment of elements with low first ionization potential (FIP effect). Here we study the spatial distribution and temporal evolution of the coronal chemical composition in an active region (AR) over about 10 days, using \hinode/EIS spectra, and we also analyze coordinated \iris\ observations of the chromospheric and transition region emission to investigate any evidence of the footprints of the FIP effect in the lower atmosphere. To derive the coronal abundances we use a spectral inversion method recently developed for the \muse\ investigation \citep{Cheung2019,Depontieu2020}. We find that in the studied active region (AR~12738) the coronal FIP bias presents significant spatial variations, with its highest values ($\sim 2.5-3.5$) in the outflow regions at the boundary of the AR, but typically modest temporal variability. Some moss regions and some regions around the AR sunspot show enhanced FIP ($\sim 2-2.5$) with respect to the AR core, which has only a small FIP bias of $\sim 1.5$. The FIP bias appears most variable in some of these moss regions. The \iris\ observations reveal that the chromospheric turbulence, as derived from \iris$^2$ inversions of the \mgii\ spectra, is enhanced in the outflow regions characterized by the high FIP bias, providing significant new constraints to both models aimed at explaining the formation of AR outflows and models of chemical fractionation.  
\end{abstract}
 
% The main body of the paper

\section{Introduction}
\label{introduction}

The chemical composition of solar plasmas is observed to vary substantially in different regions of the Sun. Early spectroscopic studies of the solar corona revealed departures from the underlying photospheric composition \citep[e.g.,][]{Meyer85,Feldman1992}. In particular, in the solar corona, elements with low First Ionization Potential (FIP) -- such as, e.g., Mg, Fe, Si -- are typically found to be more abundant (by a factor 2-4) compared with high-FIP elements -- such as, e.g., C, N, O; this phenomenon is therefore called FIP effect, and the extent of the enhancement of element abundances is called FIP bias. The apparent dependence of chemical fractionation on the first ionization potential of the elements indicates that this process is likely occurring in the chromosphere, where neutrals with low FIP are ionized. This also suggests a link to the processes responsible for coronal heating \citep[see e.g.,][and references therein]{Testa2010,Testa2015,Laming12,Laming2015}, which are still poorly understood \citep[e.g.,][]{Klimchuk2006,Testa2015,Testa_Reale_2022arXiv}.
Further studies showed that chemical fractionation varies substantially in different regions of the Sun with varying magnetic topologies (see example in \citealt{Brooks2015}).  In particular, a strong FIP effect is often observed in high temperature ($\gtrsim 2$~MK) active region (AR) core loops, post flare loops, $\sim 1$~MK AR fan loops, AR outflows, and coronal mass ejections \citep[e.g.,][]{Feldman1992,Warren11,Brooks11,Zurbuchen2016}. However, in the transition region, newly emerging ARs, coronal holes, and transient heating events such as microflares and flares, typically the chemical composition appears closer to photospheric \citep[e.g.,][]{McKenzie1992,Widing1997,Warren2016,Young2018}. Some locations in ARs and flares sometimes even show an inverse-FIP effect \citep[e.g.,][]{Doschek2015,Doschek2016,Doschek2017,Brooks2018,Baker2019,Baker2020} —i.e., with high-FIP elements enhanced relatively to low-FIP elements—, which is also typically observed in more active stars \citep[e.g.,][]{Testa2010,Testa2015}.
Furthermore, the chemical composition of the solar wind is an indicator of the source region on the Sun (e.g., \citealt{Brooks11,Brooks2015}), and is critical to establishing the magnetic connectivity from the wind to the surface.

\begin{deluxetable*}{ccccccccc}[!ht]
\label{table_obs}
\tablecaption{List of analyzed Hinode/EIS and \iris\ observations of AR~12738.
%\jms{You could add the location or mu angle and perhaps some relevant results, but not sure which ones.}
}
\tablehead{
  \multicolumn{3}{c}{Hinode/EIS} & & \multicolumn{5}{c}{\iris} \\ \cline{1-3}\cline{5-9}
 \colhead{startime [UT] } & \colhead{study name} & \colhead{$x,y$} & & \colhead{startime [UT]} & \colhead{OBSID} & \colhead{texp [s]} & \colhead{n.\ rasters} & \colhead{$x,y$}
}
\startdata
\multirow{2}{*}{2019-04-10 05:36} & \multirow{2}{*}{HPW021VEL260x512v2} & -757",158" & & 2019-04-10 05:47    & 3620259477 & 8 & 1 & -758",139" \\
  &    & & & 2019-04-10 12:15    & 3620108077 & 8 & 5 & -669",169" \\
2019-04-11 15:32  & HPW021VEL260x512v2 & -620",210" & & 2019-04-11 15:02    & 3620106077 & 4 & 3  & -605",215" \\
2019-04-12 15:18  & HPW021VEL260x512v2 & -314",146" & & 2019-04-12 14:52    &  3620108077 & 8 & 1 & -253",186" \\
2019-04-13 00:55  & HPW021VEL260x512v2 & -225",147" & & 2019-04-13 01:34    &  3620108077 & 8 & 2 & -183",227" \\
2019-04-13 10:19  & HPW021VEL260x512v2 & -154",124" & & 2019-04-13 11:39    &  3620108077 & 8 & 2 & -62",189" \\
2019-04-15 18:47  & HPW021VEL260x512v2 & 340",118" & & 2019-04-15 12:43    &  3620108077 & 8 & 2 & 368",192" \\
2019-04-15 20:24  &  HPW021VEL260x512v2 & 351",149" & & 2019-04-15 21:53    &  3620108077 & 8 & 2 & 443",201" \\
2019-04-15 22:02  & HPW021VEL260x512v2  & 361",108" & & 2019-04-15 21:53    &  3620108077 & 8 & 2 & 443",201" \\
2019-04-16 01:03  & HPW021VEL260x512v2  & 392",96" & & 2019-04-15 21:53    &   3620108077 & 8 & 3 & 443",201" \\
2019-04-16 02:40  & HPW021VEL260x512v2  & 391",114" & & 2019-04-16 02:01    &  3620108077 & 8 & 1 & 476",199" \\
2019-04-17 01:37  & HPW021VEL260x512v2  & 699",65" & & 2019-04-17 01:16    &  3630108077 & 8 & 3 & 677",173" \\
2019-04-17 16:42  & HPW021VEL260x512v2  & 782",68" & & 2019-04-17 15:42    &  3610108077 & 8 & 7 & 790",158" \\
2019-04-18 02:13  & HPW021VEL260x512v2  & 829",65" & & 2019-04-18 12:08    &   3620108077 & 8 & 4 & 886",133" \\
2019-04-19 04:25  & HPW021VEL260x512v2  & 872",85" & & 2019-04-19 04:48    &  3620108077 & 8 & 3 & 924",123" \\
2019-04-19 16:14  & HPW021VEL260x512v2  & 919",21" & & 2019-04-19 12:20    &  3620108077 & 8 & 6 & 934",117" \\
\enddata
\end{deluxetable*}

In ARs the coronal chemical composition has also been observed to change as the AR evolves, but with contradictory results. Early \skylab\ observations of newly emerged ARs point to a significant increase of FIP bias with the aging of the AR by up to an order of magnitude \citep{Widing2001}. A couple of more recent studies based on observations with the \hinode\ \citep{Kosugi2007} Extreme-ultraviolet Imaging Spectrograph (EIS; \citealt{Culhane2007}) have found examples of evolution of the chemical composition in ARs that appear at odds with the earlier \skylab\ findings. In particular, \hinode/EIS observations of AR 11389, a mature AR in its decay phase, sampling its coronal composition over a couple of days, point to a decrease of FIP bias with time \citep{Baker2015}. Another EIS spectral study of an unnumbered decaying AR show some similarities, with large fluctuations and an overall decrease of FIP bias over about 4 days of the advanced decay phase \citep{Ko2016}.  These very few studies of FIP bias evolution in ARs indicate that compositional changes in coronal plasma in AR are not as simple as suggested by early \skylab\ studies, and that the more complex relationship between abundance anomalies and AR properties and evolution needs to be determined through more extended studies. 

Recent promising models have been developed in which chemical fractionation is driven by the ponderomotive force of Alfvén waves due to the propagation and/or reflection of MHD waves in the chromosphere
\citep{Laming04,Laming09,Laming12,Laming2015}. In these models there is an intimate connection between the processes leading to chemical fractionation and those responsible for coronal heating, therefore suggesting the abundance anomalies can also yield important insights into the long-standing issue of the heating of stellar coronae. Although these models are able to generally reproduce scenarios in which FIP effect and inverse FIP effect can occur (e.g., \citealt{Laming2015}), not all observational aspects are predicted (e.g., the extent of variability of FIP effect in ARs; \citealt{Doschek2019}), and further observational constraints on these models are needed to make significant progress \citep{Laming2015}.

Robust measurements of the dependence of chemical fractionation on active region properties and evolution (magnetic field complexity and evolution, and coronal activity evolution) and of correlations between coronal composition and chromospheric/transition region (TR) properties have the potential to put very tight constraints on models, and therefore shed light on the physical processes leading to fractionation. However, the paucity of existing studies of variability of the coronal composition during a significant portion of the AR evolution has not provided a clear picture of its properties and evolution. Also, there is a lack of studies exploring the connection with the lower atmospheric layers, which are where the fractionation is expected to happen. 

In this paper we provide new constraints to the models by analyzing coordinated observations of an AR observed for over 10 days by \hinode/EIS and by the Interface Region Imaging Spectrograph (\iris, \citealt{DePontieu14}). \hinode/EIS spectra are used to derive the spatial and temporal properties of coronal abundances, while \iris\ spectral observations provide insights into the underlying chromospheric and transition region properties, allowing us to explore the correlations between coronal composition and lower atmospheric conditions.

\begin{figure*}[!ht]
	\centering
	\includegraphics[height=12cm]{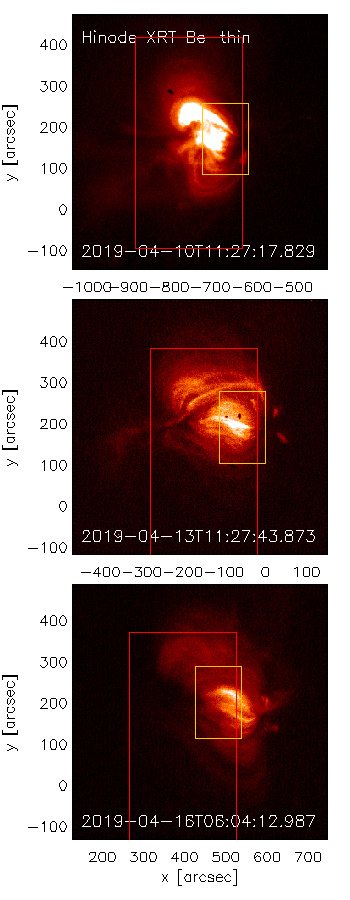}\hspace{-1cm}	
	\includegraphics[height=12cm]{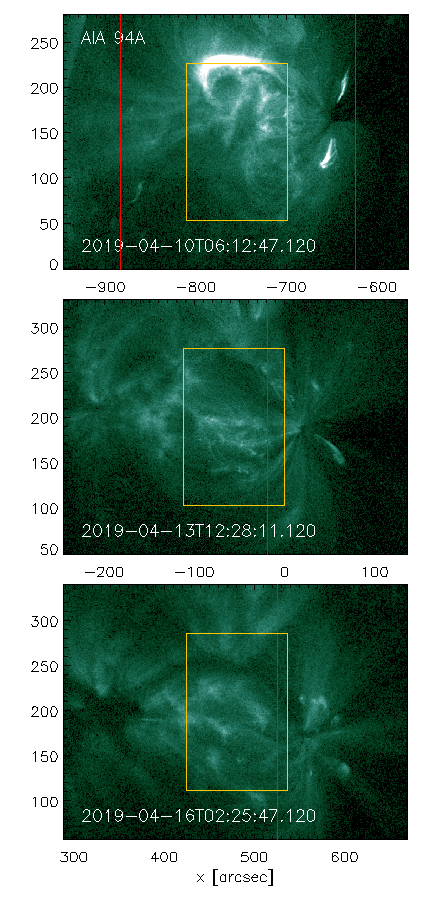}\hspace{-1cm}	
    \includegraphics[height=12cm]{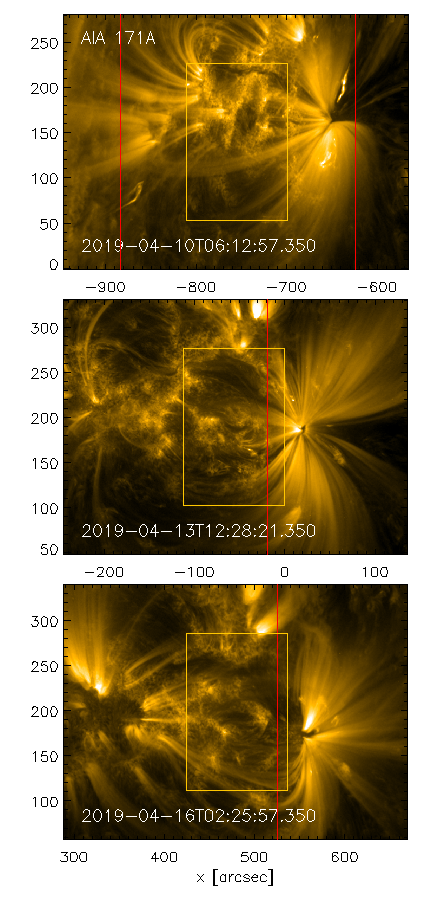}\hspace{-0.8cm}	
    \includegraphics[height=12cm]{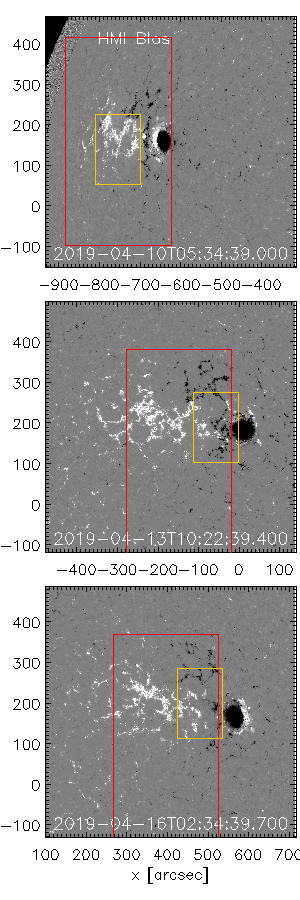}	
	\caption{\hinode\ and \sdo\ observations of AR~12738 at three different times of the AR disk passage, roughly every 3 days, to show the general morphology of the coronal emission at different temperatures and of the photospheric line-of-sight component of the magnetic field, and their evolution. From left to right we show images of: coronal emission (1) in the X-ray \hinode/XRT Be-thin passband, which show the hotter coronal emission, (2) in the AIA 94\AA\ narrowband (which has a cooler, $\sim 1$~MK, component, and a hot $\gtrsim 4$~MK component), and (3) in the AIA 171\AA\ narrowband, dominated by $\sim 1$~MK emission from \feix; we also show (4) the magnetograms from \sdo/HMI. Note that the field of view is not the same for all passbands, and in particular we show a larger f.o.v.\ for XRT and HMI, while for AIA we use the datacubes coaligned with the \iris\ observations and available from the \iris\ database. The latter have a slightly larger f.o.v.\ than the corresponding \iris\ observations.  We mark with red and yellow boxes the f.o.v.\ of EIS and the \iris\ spectrograph, respectively, for the corresponding observation closest in time to each image.
	}
	\label{fig_ar_img_bands}
\end{figure*}

\section{Observations and Data Analysis}
\label{observations}

We searched the \hinode/EIS and \iris\ databases for datasets suitable for our investigation, in particular, covering several days of an AR evolution, with $\sim$daily cadence for both spectrographs, and for which the EIS observations have sufficient spectral coverage to determine the coronal abundances. We selected EIS spectra with a large set of Fe lines (\feviii-\fexvi) able to constrain the temperature distribution and density of the coronal plasma, and also including \six\ and \sx, which are typically used to derive the FIP bias from EIS spectra \citep[e.g.,][]{Brooks11,Baker2015}. 

Here we present analysis of a timeseries of EIS and \iris\ spectra of AR~12738, observed from 2019-04-10 to 2019-04-19 (listed in Table~\ref{table_obs}).
The \hinode/EIS rasters we analyze (study acronym HPW021VEL260x512v2) use the 2\arcsec\ slit, 40~s exposure time at each of the 87 slit positions, and raster steps of 3\arcsec, therefore covering a field-of-view of about 260\arcsec\ $\times$ 512\arcsec\ (in about 1 hour); this study includes a large number of lines suitable for determining temperature, density, and abundances of active region plasma.
The \hinode/EIS datasets were reduced (to correct the raw data for dark current, cosmic rays, hot, warm, and dusty pixels and to remove instrumental effects of orbital variation, CCD detector offset, and slit tilt) using standard routines available in the \hinode/EIS branch of Solar Software \citep{SSW}.
The \iris\ observations we analyze are very large dense 320-step rasters, rebinned $2 \times 2$, with exposure time of 8~s (except for only one observation on 2019-04-10 at 15:02UT which has 4s exposures), and raster steps of 0.35\arcsec, therefore covering a field-of-view of about 112\arcsec\ $\times$ 174\arcsec\ (in about 50~min for the 8~s exposure, and $\sim27$~min for the 4~s exposure).
We use \iris\ calibrated level 2 data, which have been processed for dark current, flat field, and geometrical corrections \citep{DePontieu14}. 

\begin{figure}[ht]
	\centering
	\includegraphics[width=8.5cm]{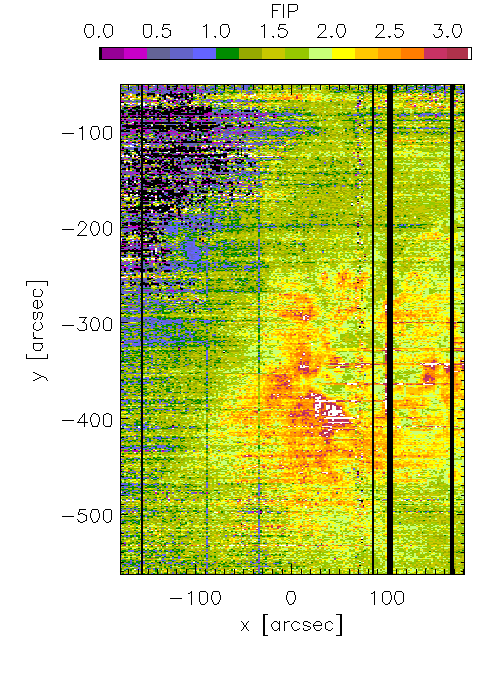}	
	\vspace{-1.5cm}		
	\caption{FIP bias map obtained by applying a modified VDEM inversion method (\citealt{Cheung2019,Depontieu2020}; see text for details) to EIS observations of AR~11389 on 2012 January 4 at 09:40 UT previously analyzed by \cite{Baker2015} using a different DEM inversion method to derive the coronal abundances. A comparison between our results shown here and their results shown in Fig.~2 (middle left panel) of \cite{Baker2015} demonstrate excellent agreement between the two methods.}
	\label{fig_baker15}
\end{figure}

We also use coordinated imaging observations, taken with the Atmospheric Imaging Assembly (AIA; \citealt{Lemen2012}) onboard the Solar Dynamics Observatory (\sdo; \citealt{Pesnell2012}), which are characterized by 0.6\arcsec\ pixels, and 12~s cadence and sample the transition region and corona across a broad temperature range \citep{Boerner2012,Boerner2014}. 
In particular, we used the AIA datacubes coordinated with and coaligned to the \iris\ datasets, which are distributed from the \iris\ search data page\footnote{https://iris.lmsal.com/search/}.
The AIA data are used for co-alignment between the instruments, and specifically particularly useful are the extreme-ultraviolet 193\AA\ narrow band images for co-alignment with the EIS coronal data, and the UV 1700\AA\ channel to coalign with the \iris\ data. The AIA coronal data are also used to investigate the coronal activity level of the AR and its variability.
Furthermore, we also use X-ray images taken with the \hinode\ X-ray Telescope (XRT; \citealt{Golub2007}). We analyze calibrated level 1 XRT  data (also normalized by the exposure time) for all the short exposure (0.5~s to 2.9~s) synoptic images in the Be-thin passband from 2019-04-10 to 2019-04-19; we focus on the short exposure data since we are focusing on the only AR on disk, which was therefore the brightest region in the images\footnote{XRT typically takes for each passband synoptics images with 2 or 3 different exposure times, which later are combined into a composite image with dark and bright regions all well exposed (level 2 XRT data; available at http://solar.physics.montana.edu/HINODE/XRT/SCIA/latest\_month.html).}. 

The EIS and \iris\ spectral data were analyzed using a variety of methods, as we briefly describe in the following. 
Most previous EIS studies aimed at determining coronal abundances (e.g., \citealt{Brooks11,Baker2013,Baker2015,Baker2018, Ko2016}) use a FIP diagnostic based on single line fits with the following analysis steps: (a) the plasma temperature distribution (also called differential emission measure, DEM) is derived using a set of Fe lines formed at different temperature (\feviii-\fexvi), and taking into account the sensitivity of the line emissivities by using densities diagnosed from \fexiii\ density-sensitive line pairs, and then (b) measured ratios of the \six\ 258.38\AA\ to \sx\ 264.23\AA\ line, are compared to the values predicted using the DEM obtained in step \textit{(a)} to derive the FIP bias (the underlying assumption being that Si and Fe, both low FIP elements, vary in a similar way).
These previous EIS studies reconstruct the DEM by using a Markov-Chain Monte Carlo algorithm \citep{Kashyap98} which has the advantage of providing estimates of uncertainties for the derived DEMs, but has the disadvantage of being markedly slow (this likely being one of the main reasons for the paucity of extensive abundance studies analyzing the temporal variation in large fields of view). 

\begin{figure*}[!ht]
	\centering
	\includegraphics[width=18cm]{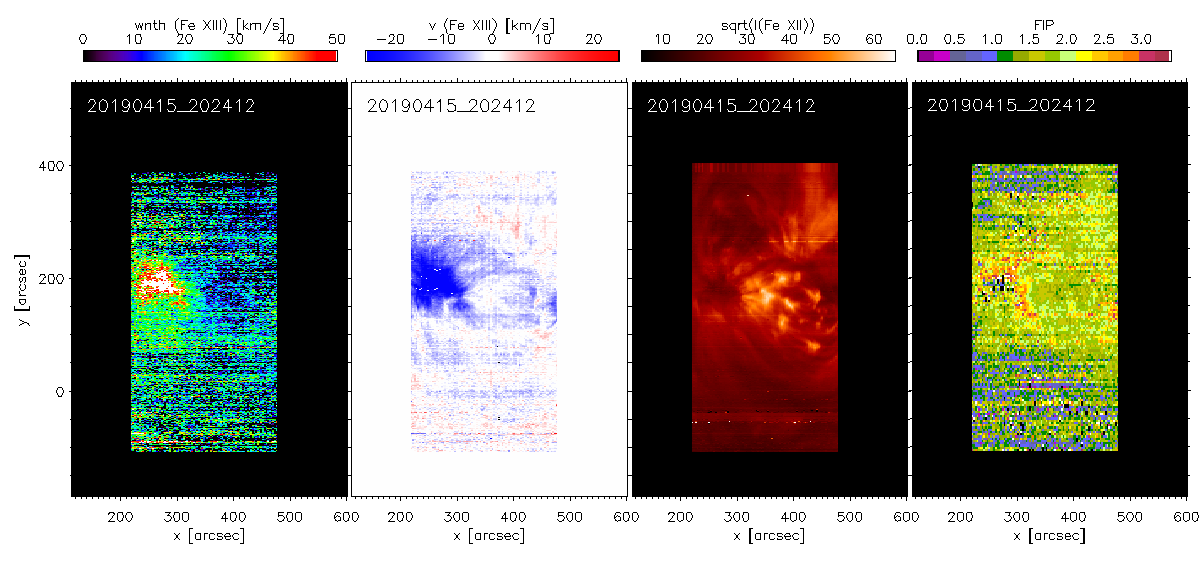}	
	%\vspace{-1.5cm}		
	\caption{Example of maps of coronal plasma properties derived from the \hinode/EIS observations, for one of the 15 EIS datasets analyzed (see also Figure~\ref{fig_ar_fip_time}), when AR~12738 was close to disk center. From left: non-thermal velocity and Doppler shift from the \fexiii\ 202\AA\ line, intensity in the \fexii\ 195\AA\ line, and  FIP bias, derived as described in section~\ref{observations}.}
	\label{fig_ar_dc}
\end{figure*}

Analogously to the above cited EIS studies, we used single Gaussian functions to fit the unblended calibrated spectra for \feviii\ (185.21\AA), \feix\ (188.50\AA), \fex\ (184.54\AA), \fexiv\ (264.79\AA), \fexv\ (284.16\AA), \fexvi\ (262.98\AA), \sx\ (264.23\AA), and \six\ (258.38\AA), while multiple components were used to fit the blended regions including \fexi\ (188.22\AA), \fexii\ (195.12\AA), and \fexiii\ (203.83\AA). Here however we apply a new inversion method, which is much faster (more than an order of magnitude), while retaining significant accuracy, as we will show in the following. This method is based on a compressed sensing method \citep{Cheung2019} we developed for the future NASA MIDEX mission \muse\ (Multi-slit Solar Explorer; \citealt{Depontieu2020,DePontieu2022,Cheung2022}) to robustly retrieve plasma properties (such as DEM, plasma velocity, density) from a variety of possible spectrometer configurations. In \citet{Cheung2019} one of the several tests of the method showed its application to EIS data (their Fig.~3).
This technique has recently also been applied to derive the magnetic field from magnetic-field-induced transitions (MIT) using 3D radiative MHD numerical models \citep{MartinezSykora2022}. 
Here we modified the method to derive DEM, density, and abundances from the set of EIS lines. A paper in preparation (Martinez-Sykora \& Testa, in prep.) will present a detailed description of the method and the tests we performed also using 3D MHD numerical simulations (similar to what was presented in \citealt{MartinezSykora2022}).
As a demonstration of the performance of this new method, we have applied it to EIS spectral observations of AR~11389 on 2012 January 4 at 09:40 UT previously analyzed by \cite{Baker2015}, to compare our results with theirs. The FIP bias map we obtained is shown in Figure~\ref{fig_baker15}, and is in excellent agreement with the map \cite{Baker2015} presented in their Fig.~2.

The analysis of \iris\ spectra includes the analysis of the TR and chromospheric emission. We fit the \siiv\ 1402\AA\ line using a single Gaussian function to derive line intensity, Doppler shift, and non-thermal broadening. We also apply \iris$^2$ inversions to the optically thick Mg II h \& k chromospheric lines: this inversion method is based on machine and deep learning techniques that allow the inference of the thermodynamic conditions of the lower atmosphere from \iris\ spectra by taking into account non-LTE conditions in the chromosphere \citep{SainzDalda2019}.

\section{Results}
\label{sec:res}

\begin{figure*}[!ht]
	\centering
	\includegraphics[width=5.8cm]{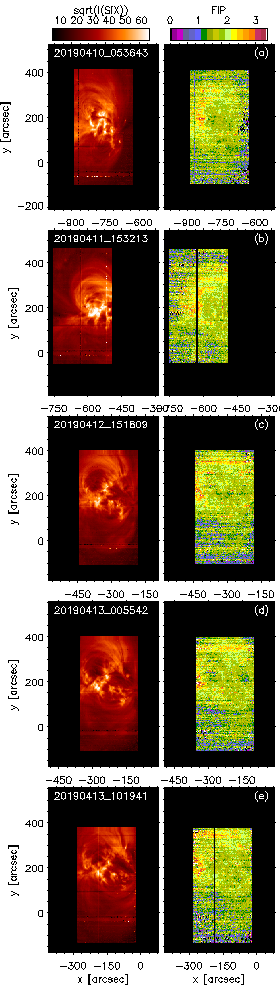}	
	\includegraphics[width=5.8cm]{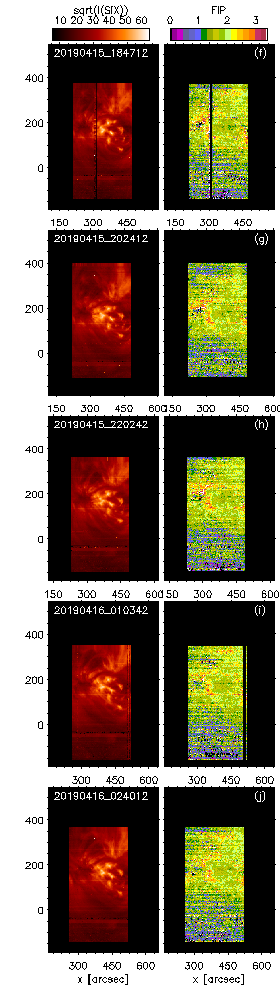}	
    \includegraphics[width=5.8cm]{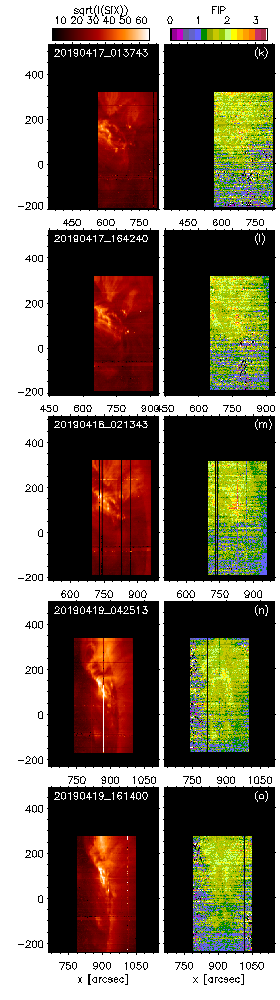}	
	%\vspace{-1.5cm}		
	\caption{Time series of maps of FIP bias for the 15 \hinode/EIS observations of AR~12738 in the $\sim 10$~days period between 2019-04-10 and 2019-04-19. For each observation we also display the emission in the \six\ line to show the morphology of the coronal emission at $\sim 1.5$~MK for the different times and line-of-sights sampled by the EIS observations. The EIS maps are shifted so that the center of the f.o.v.\ corresponds to the AR location at the center of the f.o.v.\ in the first observations, with the solar rotation appropriately taken into account at each time.
	}
	\label{fig_ar_fip_time}
\end{figure*}

The objectives of our study are (a) to analyze the spatial distribution and temporal evolution of the coronal abundance anomalies in an active region observed for a  significant portion of its limb-to-limb passage, and (b) to investigate whether a footprint of the chemical fractionation can be discerned in the lower atmosphere (chromosphere and transition region). 
We first describe our findings related to the former objective and then we discuss our exploration relevant to the latter objective.

\subsection{FIP spatial and temporal distribution}\label{subs:FIP}
As described in the previous section, we use a dozen  lines, including Fe lines of different ionization stages, providing temperature and density diagnostics, and \sx\ and \six\ lines that constrain the FIP bias. The analysis provides for each observation a map of several quantities: line intensities, Doppler shift, and non-thermal broadening can be derived from the line fitting, and the inversion method additionally provides FIP bias, and emission measure as a function of density and temperature (DEM(n,T)).  For the DEM inversion we assumed a density grid with 0.3 bin size in Log(n), between 8 and 11, while for the temperature we assume Log(T) binning of 0.05, between 5.5 and 6.75. Note also that the \sx\ is a weak line, and it has relatively low signal-to-noise in several areas of the active region, and therefore we have rebinned the \sx\ (and \six) spectra by a factor 4 in the y direction, so the FIP bias maps have effective pixels of 2\arcsec $\times$ 4\arcsec. 

\begin{figure*}[!ht]
	\centering
	\includegraphics[width=18cm]{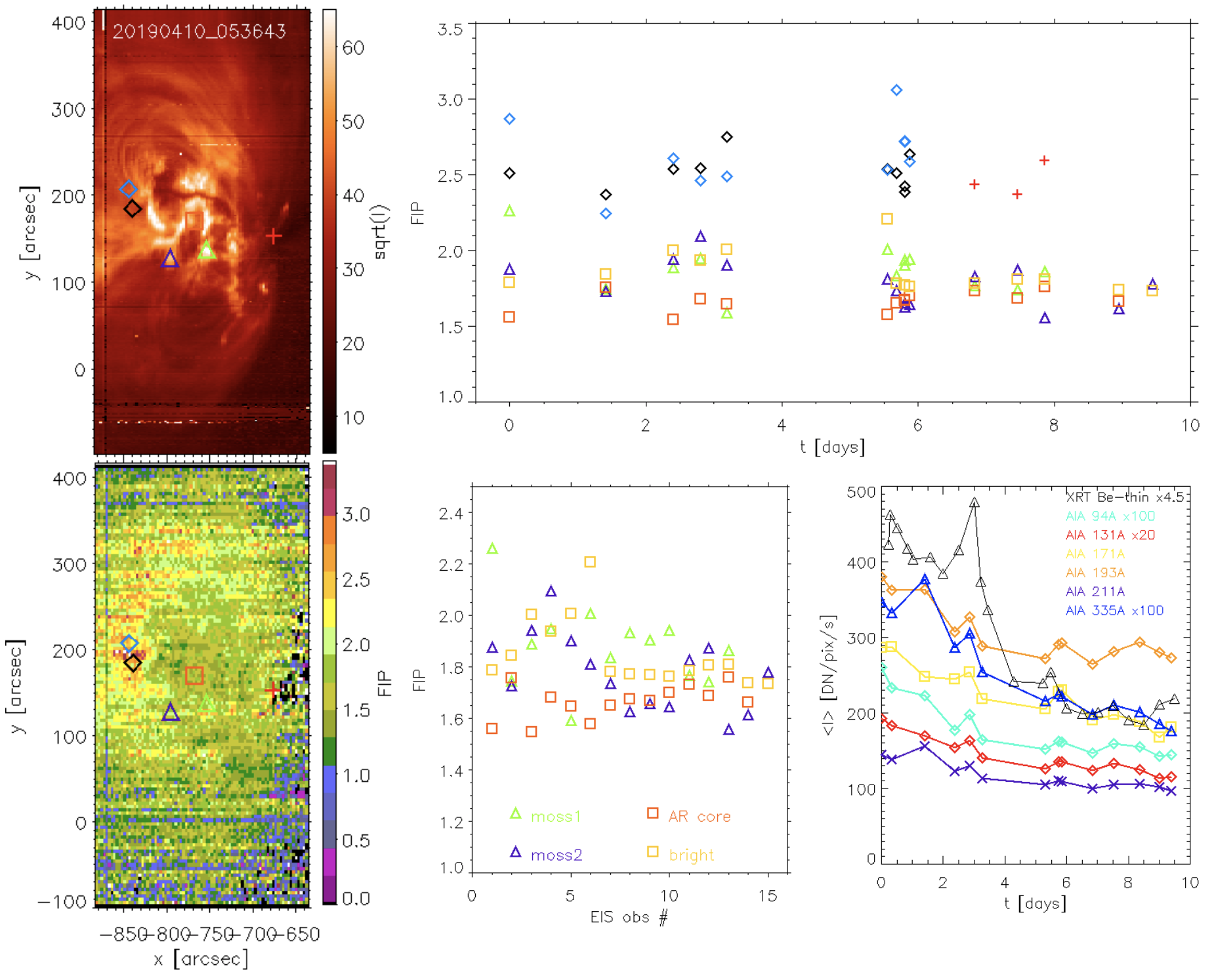}
	\caption{Temporal variability of coronal properties of AR~12738 from \hinode\ and \sdo/AIA observations over about 10~days. The temporal variability of the FIP bias is shown ({\em right top panel}) for a few selected locations marked with corresponding symbols in the left panels, where they are superimposed on the maps of coronal emission (top) and FIP bias (bottom) for the first \hinode/EIS observation. We chose two locations in the outflow region (diamonds), two in moss regions (triangles), and one close to the sunspot in the western side of the AR (pluses), as well as one in the AR core (red square). The FIP values for each location are averaged over an area of about 14\arcsec$\times 14$\arcsec\ (and double that for the AR core, marked with the red square). The yellow-orange squares represent averages of FIP over the brightest regions in the AR in \six\ emission (we use the same threshold of intensity for all observations). Not all locations are in the \hinode/EIS f.o.v.\ at all times due to the variation in pointings, which focus on different parts of the AR at different times (see also Figure~\ref{fig_ar_fip_time}). In the middle panel of the bottom row, we show a subset of the results in the top right panel, focusing on the moss, AR core, and bright regions; in this plot we use observation number for the x-axis to better resolve observations too close in time, for which the symbols overlap in the top panel (at $t\sim6$ day). In the bottom right panel we show lightcurves of coronal emission in a \hinode/XRT bands and the 6 AIA coronal passbands (see text for detail). }
	\label{fig_fip_temp}
\end{figure*}

In Figure~\ref{fig_ar_dc} we show an example of the maps of the line properties (intensity, Doppler shift, and non-thermal broadening) as well as of the FIP bias, obtained from an observation on 2019-04-15, when the AR~12738 was not far from disk center. The Doppler shift and non-thermal broadening (from \fexiii) spatial distribution is in agreement with what is typically found from \hinode/EIS observations in active regions, with large blueshifts and broadening (for $Log(T[K]) \gtrsim 6.5$) at the weakly emitting edges of AR where the so-called ``AR outflows" are. The velocities and broadening are otherwise typically small in the AR (see also, e.g., \citealt{Brooks11,Brooks2016}). The FIP bias is also increased in the outflow regions, in agreement with previous studies (e.g., \citealt{Brooks11,Brooks2015}),  which pointed out this property as a potential way to trace sources of parts of the slow solar wind to the solar surface. Most of the AR plasma shows a modest FIP bias of $\sim 1.5$ to 2. 

Figure~\ref{fig_ar_fip_time} presents the maps of FIP bias, as well as coronal morphology (\six\ intensity), for the full timeseries of EIS observations analyzed here, from close to the eastern limb (on 2019-04-10) to the West limb (on 2019-04-19; see Table~\ref{table_obs}).
These data indicate that a large FIP bias is typically found in the outflow regions (see eastern edge of AR, which is in EIS f.o.v.\ for the observations on 04-10 to 04-16), in some large and cool (fan) loops (at the northern side of the EIS f.o.v., especially on 04/10-12), in some moss regions (the moss being the high density TR of hot cool loops; e.g., \citealt{Fletcher1999,Brooks2009,Tripathi2010,Testa2013}; see bright \six\ structures in the AR core),
and, at times, close to the sunspot  (on western side of AR, in EIS f.o.v.\ for the observations on 04-15 to 04-17). 

Very limited evolution is observed in the FIP bias of outflow regions, where it is consistently high ($\sim 3$), and in the AR core, where it is $\lesssim 2$.
Some moss regions are where the FIP appears to be most variable (over timescales of hours, to which the cadence of these EIS observations is sensitive to), with some of these regions intermittently showing enhanced FIP with respect to the rest of the AR core.  
To illustrate these results, in Figure~\ref{fig_fip_temp} we show the variability of FIP bias values for a few different locations sampling outflow regions, moss, and the AR core.  We average the FIP values for each location over an area of about 14\arcsec$\times 14$\arcsec\ (and double that for the AR core). We note that not all locations are in the \hinode/EIS f.o.v.\ at all times, because of the variation in pointings, which focus on different parts of the AR at different times (see also Figure~\ref{fig_ar_fip_time}).
We also derived the average FIP for the brightest regions in the \six\ emission (yellow-orange squares in Figure~\ref{fig_fip_temp}). The FIP in these bright regions shows some variability, in particular with an increase from initial values of $\sim 1.8$ to $\sim 2.3$ in the first $\sim 6$ days of the time series, and then a decrease to $\sim 1.7$. These brightest regions are interesting also for comparison with previous works of \cite{Baker2015} and \cite{Ko2016} which mostly focused on the bright AR core, although we note that part of the variation we observe might be due to the changing EIS f.o.v.\ which covers different portions of the AR at different times (see Fig.~\ref{fig_ar_fip_time}). 

In Figure~\ref{fig_fip_temp} we also show lightcurves in several coronal passbands observed with \hinode/XRT and \sdo/AIA. For XRT we selected all the short-exposure synoptic images in the Be-thin passband, in which the AR core is bright (but not saturated), and we average the intensity in the f.o.v.\ shown in Figure~\ref{fig_ar_img_bands} (left panel). For the AIA coronal narrowbands we use the AIA datacubes coaligned to the \iris\ data (see middle panels in Figure~\ref{fig_ar_img_bands}), and average the emission in the whole f.o.v.\ and for the entire timeseries ($>1$~hr). The comparison of these lightcurves with the plots of the FIP variability in various locations does not show a clear correlation between the FIP and the coronal properties, except possibly for the average in \six\ bright regions (yellow-orange squares in Figure~\ref{fig_fip_temp}) which is lower in the last 4 days when the coronal activity has decreased (although the peak is around day 5.5, when the decrease in coronal activity had already happened).

\begin{figure*}[!ht]
	\centering
	%\hspace{-0.5cm}		
	\includegraphics[height=22cm]{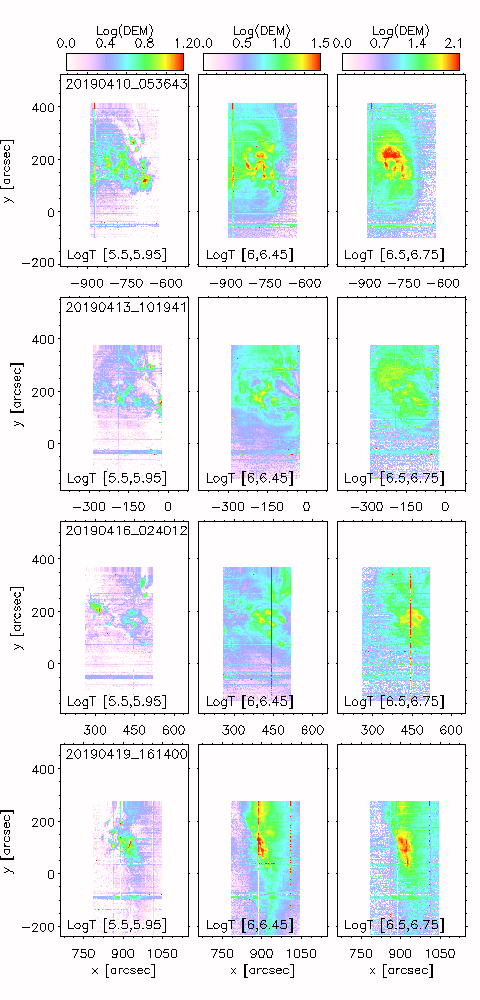}	
	\includegraphics[height=22cm]{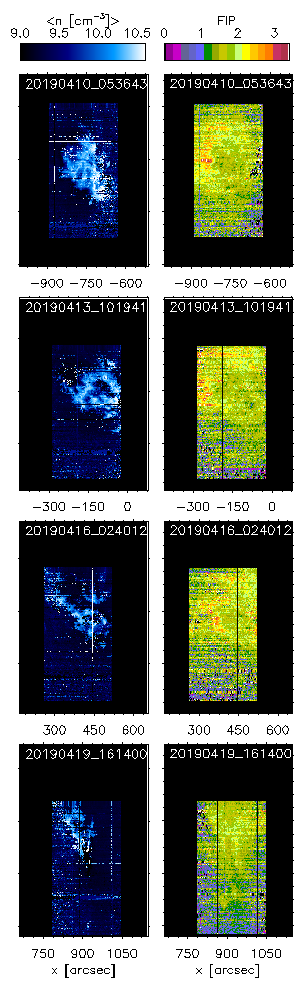}	
	\vspace{-0.5cm}		
	\caption{Evolution of DEM in 3 temperature bands (for Log(T[K]) below 6, between 6 and 6.5, and 6.5 and above, first 3 columns, from left to right respectively), of plasma density (fourth column) derived as the average of density weighted by the DEM(n,T), and of FIP (rightmost column), for four datasets sampling (every $\sim 3$~days) the entire interval of the AR observations analyzed in this paper.}
	\label{fig_ar_dem_time}
\end{figure*}

The analysis of EIS spectra also provides information on the plasma density and thermal distribution in the active region, and their evolution. Although they are not a primary focus of this work, we summarize the results in Figure~\ref{fig_ar_dem_time}. We show, for four datasets sampling the EIS timeseries every $\sim 3$~days, the maps of emission measure integrated in three different temperature ranges (and over all density values), and the maps of the density obtained as a weighted average, using DEM(n,T) as weights. Specifically, we define the weighted average density in the spatial pixel of coordinate $(x,y)$ as:
\begin{equation}
    <n_{x,y}>=\frac{\sum_n n \times (\sum_T DEM(x,y,n,T))}{\sum_n \sum_T DEM(x,y,n,T)}
\end{equation}
where $DEM(x,y,n,T)$ is the DEM, as a function of density and temperature, in that spatial pixel.
This figure shows that the active region is not very active, and it is characterized by a small decrease in density and emission measure after the first few days. Indeed, AR~12738 was the only AR on disk for most of the timeseries analyzed here, and we can use the GOES X-ray curve as a proxy of its activity: the observed GOES X-ray level is always at or below B level during the $\sim 10$~days interval of our observations, with a few mid-B-class events observed in the first few days (until 2019-04-14) and no events from 04-15 to 04-19. Similarly, the \hinode/XRT and AIA lightcurves we showed in Figure~\ref{fig_fip_temp} indicate an overall decrease of activity after the first three days.

\begin{figure*}[!ht]
	\centering
	%\hspace{-0.5cm}		
	\includegraphics[width=18cm]{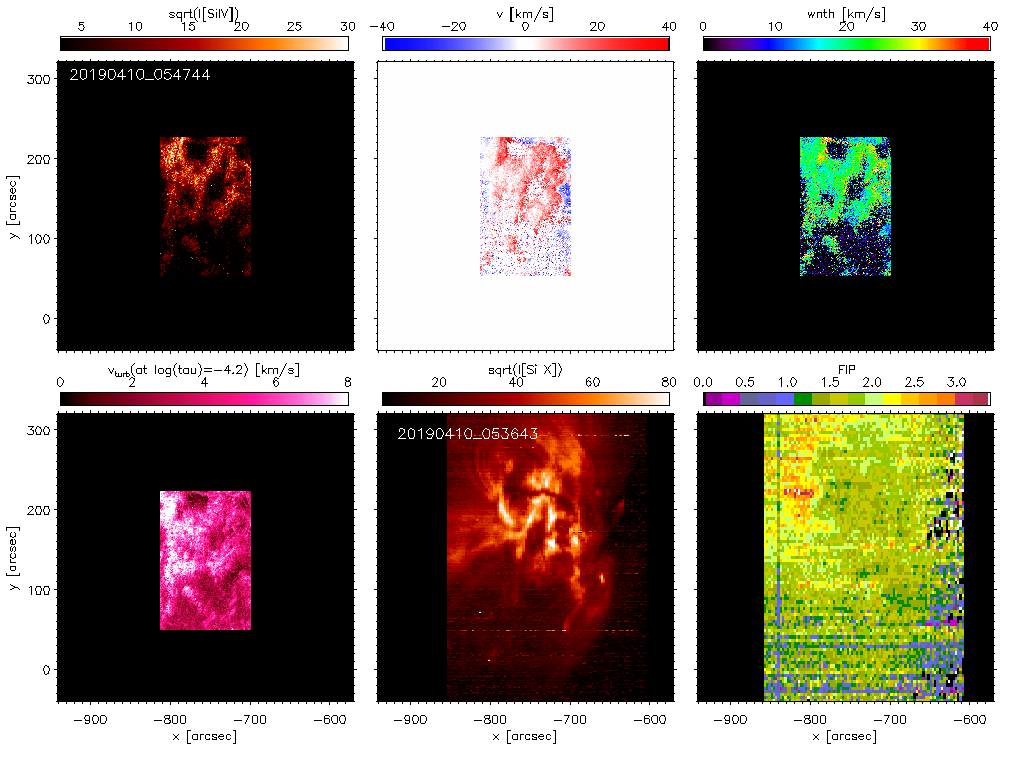}	
	%\vspace{-1.5cm}		
	\caption{Example of maps of chromospheric, transition region, and coronal properties from the first of the coordinated observations: \iris\ \siiv\ intensity, Doppler and non-thermal velocity ({\em top row}), chromospheric turbulence (at $\tau = -4.2$) derived from \iris$^2$ inversions of the \mgii\ spectra ({\em bottom left}), and coronal emission and FIP bias ({\em bottom} middle and right panel respectively). The areas with enhanced FIP bias (mostly in the eastern side of the AR) also appear to have increased chromospheric turbulence in the areas of overlap of \iris\ and EIS f.o.v.\ (with a Pearson correlation coefficient of 0.35 for this dataset).}
	\label{fig_ar_iris2}
\end{figure*}

\subsection{\iris\ chromospheric and transition region properties, and relations with coronal FIP}\label{subs:IRIS}

\begin{figure*}[!ht]
	\centering
	%\hspace{-0.5cm}		
	\includegraphics[width=8cm]{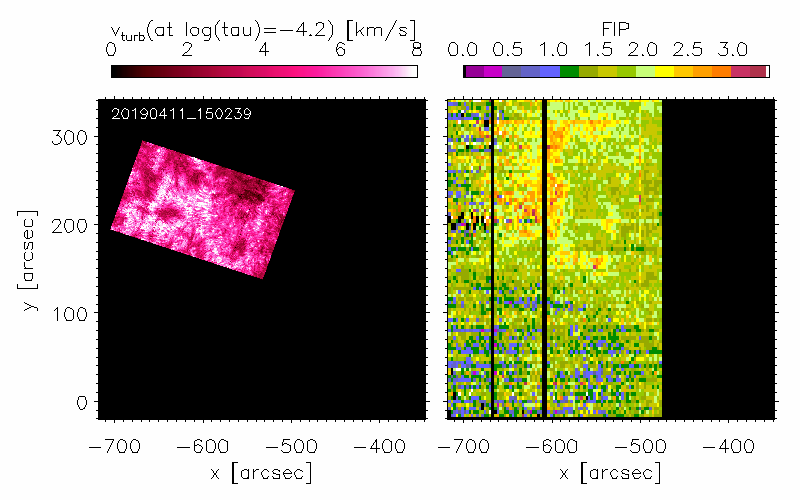}	
	\includegraphics[width=8cm]{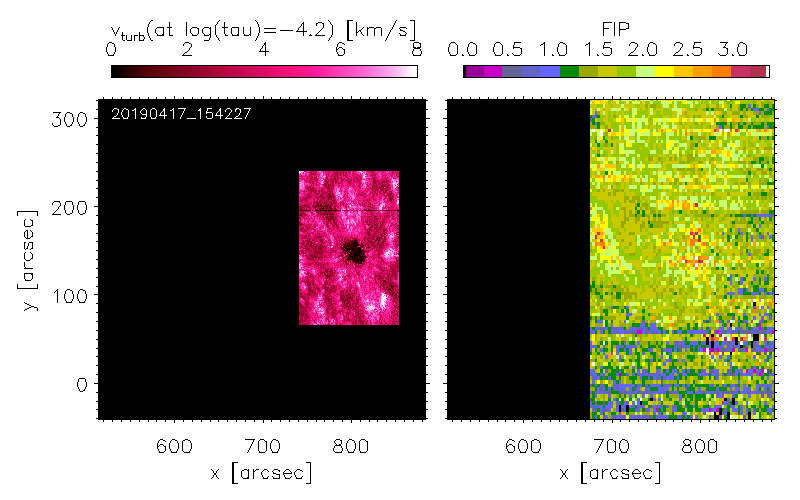}\vspace{-0.2cm}	
	\includegraphics[width=9cm]{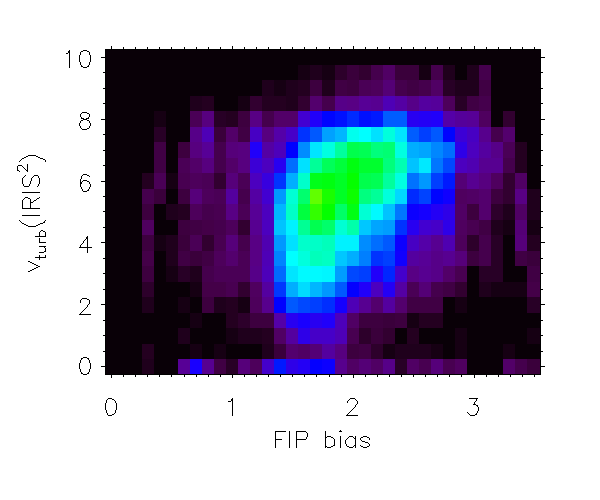}
	\hspace{-1cm}
	\includegraphics[width=9cm]{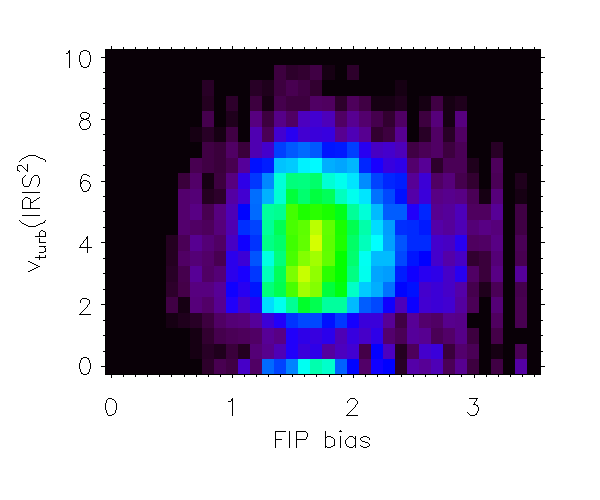}\vspace{-0.8cm}
	%\vspace{-1.5cm}		
	\caption{For two more coordinated \iris\ and EIS datasets we show the maps of chromospheric turbulence and FIP bias ({\em top row}) and the 2D histograms ({\em bottom row}) in the space of these two parameters. A moderate correlation (0.31 in this case, comparable with the 0.35 of the first dataset of Figure~\ref{fig_ar_iris2}) is found when the high-FIP bias regions in the eastern side of the AR (largely corresponding to the outflow regions) are within the f.o.v.\ of both instruments, whereas no significant correlation is observed otherwise.}
	\label{fig_fip_vturb}
\end{figure*}

We then analyzed the coordinated \iris\ observations and derived for each dataset maps of \siiv\ line properties (intensity, velocity, and non-thermal broadening), and applied \iris$^2$ inversions to the \mgii\ spectra to derive a model atmosphere (chromospheric temperature, electron density, line-of-sight velocity, $v_{los}$, and turbulent velocity, $v_{turb}$ --also known as microturbulence--, as a function of the optical depth $\tau$ at 500 nm). 
In Figure~\ref{fig_ar_iris2}, we show an example of these maps from the start of the observing sequences on 2019-04-10, where for the \iris$^2$ results we show the map of $v_{turb}$ at $\tau =-4.2$ (typically the inversions are better constrained in the $\tau$ range $\sim -3.8$ to $\sim -5$). The choice of showing the $v_{turb}$ is motivated by both theoretical and observational results: the fractionation model of Laming (e.g., \citealt{Laming2015} and reference therein) predicts a dependence of FIP on the magnetic waves propagating and/or reflecting in the chromosphere, and recent observations of a sunspot found Alfv\'enic waves associated with FIP enhancements \citep{Stangalini2021,Murabito2021,Baker2021}. Assuming all the microturbulence comes from temporal unresolved \alfven\ waves, the time integration provides an upper limit of the frequencies, and the values of the microturbulence will be associated with the amplitude. However, other physical processes may contribute to micro-turbulence, e.g., other unresolved flows, jets, turbulence, heating, or opacities. 
Furthermore, a visual inspection of the \iris$^2$ derived quantities indicates a potential correlation of $v_{turb}$ around $\tau =-4.2$ with the FIP as visible in Figure~\ref{fig_ar_iris2}. For the observation shown in Figure~\ref{fig_ar_iris2}, the Pearson cross-correlation coefficient between FIP and $v_{turb}$ at $\tau =-4.2$ is 0.35, indicating a moderate correlation between the two variables. We find that the correspondence between FIP and $v_{turb}$ is mostly evident in the areas in the eastern side at the boundary of the AR, particularly in the outflow regions. 
The presence of correlation is interesting and not necessarily expected, given the connectivity between the high-FIP coronal regions and the footpoints which might not overlap, and therefore the observed correlation might be underestimated.
We note that calculations of the cross-correlation just in these high-FIP areas do not show significant correlation, and that this is not unexpected, also because at any given location the \iris\ and EIS data are typically obtained at quite different times (from minutes to hours; even for EIS and \iris\ scans taken roughly simultaneously, because of the different scanning direction for the two instruments-- W-E for EIS and E-W for \iris; see Fig.~9 and text of \citealt{Testa2016} for an example). 
In Figure~\ref{fig_fip_vturb} we show two more examples of the FIP-$v_{turb}$ correlation for two observations at different times (2019-04-11 at 15UT and 2019-04-17 at 15UT). The f.o.v.\ of the EIS and \iris\ observations on 2019-04-11 covers the high-FIP area at the AR boundary, and shows a cross-correlation coefficient (0.31) similar to the first observation, whereas the latter observation has a different f.o.v.\ excluding most of those areas and no significant correlation between chromospheric $v_{turb}$ and FIP is observed.

The high FIP areas in the AR outflow regions also appear to have less redshifted \siiv\ with respect to the AR core, which generally shows significant redshift in its \siiv\ emission; these results are in agreement with our recent investigation of the TR and chromospheric counterparts of outflow regions \citep{Polito2020}. We do not see significantly different \siiv\ non-thermal broadening in the high FIP areas. This is analogous to the findings of \citet{Barczynski2021} that the measured transition region (\iris\ \siiv) non-thermal velocities are similar in upflow regions and active region cores. This is of course a very intriguing result, as other TR and chromospheric line properties, such as e.g., \siiv\ and \cii\ Doppler shifts and \mgii\ line asymmetries \citep{Polito2020,Barczynski2021}, instead have different values in outflows regions with respect to AR cores. We will speculate more on these findings in the discussion section here below.

\section{Discussion and conclusions}
\label{sec:discussion}

The composition of the solar corona and the solar wind often differs from that of the solar photosphere, typically with a relative enrichment of elements with low first ionization potential (FIP effect). This chemical fractionation is poorly understood but it can provide crucial clues about the physical processes at work in the solar atmosphere. In fact, the fractionation is surely originating in the chromosphere where the first ionization occurs, and linked to the coronal heating mechanism, but it also carries its signature throughout the many layers and magnetic structuring of the corona and into the solar wind. The variation of the chemical fractionation, both in space and in time, can therefore be used as a tracer of the mass and energy flow throughout the solar atmosphere, and provide insights on the drivers of the observed outflows. 

In this paper we have analyzed spectroscopic observations of an active region, over several days, with \iris\ and \hinode/EIS. The \iris\ and \hinode\ observations allowed us to study the solar atmosphere from the chromosphere to the transition region (TR) and the corona, and to measure the FIP effect and its evolution over several days, with about daily cadence. 
By combining observations of different atmospheric layers, we have also investigated possible correlations between the chemical fractionation 
observed in the corona with plasma properties in the lower atmosphere.

In order to derive the coronal abundance from the \hinode/EIS spectra we apply a modified version of a recent compressed sensing method for the spectral inversions (\citealt{Cheung2019}, and Martinez-Sykora \& Testa in preparation), similar to the version used by \citet{MartinezSykora2022} to derive DEM(n,T,B) using magnetic-field-induced transitions. This inversion method allows to derive the plasma temperature distribution, density, and abundances in a robust fashion and is significantly faster than the typically used MCMC method \citep{Kashyap98}, facilitating the analysis of long time series. 
Here we showed (Figure~\ref{fig_baker15}) the satisfactory results of a test of this new method against literature results \citep{Baker2015} obtained with the MCMC DEM inversion method, although a thorough presentation of the method and its testing is deferred to an upcoming paper (Martinez-Sykora \& Testa in prep.). 

We then presented the results of the application of the method to a time series of 15 \hinode/EIS observations of AR~12378 covering about 10~days.
The FIP bias maps show that: (a) most of the AR plasma shows a modest, and fairly constant, FIP bias of $\sim 1.5$ to 2; (b) high intensity (in \six\ emission) regions show slightly larger FIP bias of $\sim 1.7$-2.2, with smaller values later in the time series, as discussed in more detail below; (c) outflow regions at the AR boundary consistently show high FIP bias; (d) high FIP is also found in some large and cool (fan) loops, in some moss regions, and, at times, in some areas close to the sunspot; (e) the FIP in some moss regions changes quite rapidly (over timescales of hours), and it is intermittently found at enhanced values of $>2$ or at lower values similar to the rest of the AR core.
The time series of maps of FIP-bias obtained for these observations do not show overall very marked temporal variability in the FIP level of different solar features. Previous studies of the evolution of FIP-bias in AR indicated an increase in FIP during the initial flux emergence phase \citep{Sheeley1995,Widing1997,Widing2001,Baker2018}. For decaying ARs recent EIS studies suggest a small decrease of FIP \citep{Baker2015,Ko2016}, in contrast with analyses of \skylab\ data that seem to indicate a continued increase of FIP bias in old ARs to very large ($\gtrsim 7$) values \citep{Widing2001}. Our observations of AR~12378 show a rather quiet AR, that is slowly decaying (see Figures~\ref{fig_fip_temp} and \ref{fig_ar_dem_time}).  \cite{Baker2015} measured variations in FIP from observations 2 days apart, while \cite{Ko2016} used timeseries covering about 4 days of the AR evolution, and in both cases they observe a small FIP decrease of the order of 10\%-15\%, and in \cite{Ko2016}, where the high cadence observations sample small timescales, they find FIP variability (both increases and decreases of up to $\sim$15\%) on timescales of hours (see their Figure~10). We note that both \cite{Baker2015} and \cite{Ko2016} mostly focused on the brigth AR core, so, for a more meaningful comparison here we also derived the FIP for the brightest regions (see yellow-orange squares in Figure~\ref{fig_ar_fip_time}) and possibly see a small decay in FIP value over the 10~day period ($< 10$\%). The lower values are found in the later part of the time series when the AR is less active.

A novel aspect of our study, compared to previous FIP studies, is that we accompanied the \hinode/EIS chemical fractionation measurement with coodinated \iris\ spectral observations of the underlying chromosphere and transition region. The aim of this combined approach is to try to connect coronal abundance anomalies to conditions in the lower atmosphere, and especially the chromosphere, where the FIP effect is thought to originate. Overall, we did not find significant correlations between FIP and transition region properties, as observed by \iris\ in \siiv, besides the fact that outflow regions are characterized by relatively smaller redshifts than the typical AR core \siiv\ Doppler shifts, in agreement with previous findings by \citet{Polito2020} and \citet{Barczynski2021}. 
The \iris\ chromospheric inversions suggest a correspondence between chromospheric turbulence $v_{turb}$ and FIP, at least for the high FIP areas corresponding to outflow region. This result supports recent findings of a correlation between coronal outflows and flows in the TR and chromosphere \citep{Polito2020}. However no increase in $v_{turb}$ is evident for the high FIP areas observed close to a sunspot. 
The difference in underlying chromospheric turbulence between the high FIP areas in outflow regions and close to the sunspot might be explained by different scenarios: it might suggest differences in the mechanisms leading to the fractionation in the two types of regions, or the $v_{turb}$ in outflows might possibly be related to the formation of the flows and not be causally connected to the chemical fractionation, or it might be due to differences in viewing angle between line-of-sight and magnetic field between the two regions.

Another puzzling finding is the lack of correlation between FIP and the \siiv\ non-thermal broadening. If the chromospheric turbulence is indeed at least partly connected to \alfven\ waves, which are in some models central to the fractionation process \citep[e.g.,][]{Laming2015}, one would expect to see a trace in the transition region plasmas, specifically in the line broadening. The lack of correlation between FIP and non-thermal broadening might constrain the properties of the waves and where and how they might get dissipated.
These findings therefore provide new challenging observational constraints on modeling and theory of chemical fractionation.

In future work we will apply our new inversion method to other timeseries of \hinode/EIS and \iris\ coordinated AR observations, for ARs at different activity level and evolutionary stage to test whether the correlations between $v_{turb}$ and FIP are present is all ARs and with similar properties, and we will also investigate whether any other correlation with TR and/or chromospheric variables might be present in other ARs.

\acknowledgements
PT and JMS were funded for this work by the NASA Heliophysics Guest Investigator grant 80NSSC21K0737, and by the NASA Heliophysics Supporting Research grant 80NSSC21K1684. PT was also supported by contracts 8100002705 (\iris), and NASA contract NNM07AB07C (\hinode/XRT) to the Smithsonian Astrophysical Observatory. BDP and JMS were supported by NASA contract NNG09FA40C (IRIS). 
We are very grateful to Lucas Guliano for helping with the processing of the \hinode/XRT data.
This research has made use of NASA's Astrophysics Data System and of the SolarSoft package for IDL.
\hinode\ is a Japanese mission developed and launched by ISAS/JAXA, with NAOJ as a domestic partner and NASA and STFC (UK) as international partners. It is operated by these agencies in cooperation with the ESA and NSC (Norway).
\sdo\ data were obtained courtesy of NASA/\sdo\ and the AIA and HMI science teams.
\iris\ is a NASA small explorer mission developed and operated by LMSAL with mission operations executed at NASA Ames Research Center and major contributions to downlink communications funded by ESA and the Norwegian Space Centre.

 % The bibliography
%\bibliography{bib_stars}

\end{document}